\documentstyle[12pt]{article}
\input epsf.tex
\begin{document}

\newcommand{\DS}{\renewcommand{\baselinestretch}{2} \tiny \normalsize}
\newcommand{\HS}{\renewcommand{\baselinestretch}{1.5} \tiny \normalsize}
\newcommand{\SMS}{\renewcommand{\baselinestretch}{1} \tiny \normalsize}
\newlength{\captsize}          \let\captsize=\small
\newlength{\captwidth}         \setlength{\captwidth}{.8\textwidth}
\newlength{\beforetableskip}   \setlength{\beforetableskip}{-1.5\baselineskip}
\newcommand{\capt}[1]{\begin{center} \begin{minipage}{\captwidth}
         \let\normalsize=\captsize
         \vspace{\beforetableskip}
         \SMS
         \caption[#1]{#1}
         \DS
         \end{minipage}\\ \end{center}}

\baselineskip=12pt
\parskip=9pt
\begin{center}
{\LARGE\bf Simulation of an Abelian Higgs theory in world line (polymer)
representation.}\\ \vskip0.1in
A. Pap and P. Suranyi\\
Department of Physics, University of Cincinnati\\
\vskip0.1in
June 30, 1998\\
\vskip0.2in

\end{center}
In World Line Path
Integral representation Abelian Higgs theories admit a curvature term in
the Hamiltonian. Using Monte Carlo simulations we
show that the curvature term drives a  phase transition at
sufficiently strong coupling. In  quenched approximation this phase
transition is smooth, with implications that the model reduces to a Goldstone
theory. Critical properties of the system near the
 transition are investigated and found to be consistent with mean field
behavior.

\section{Introduction}In most field theoretic calculations, and  in
almost all numerical simulations of field theories  representations in
terms of functional integrals over field variables are used.
Yet, for a range of
theories  another representation, based on the Schwinger proper time formalism,
exists and can be utilized for simulations.  As Feynman pointed out a long time
ago~\cite{feynman} an alternative, quantum mechanical representation of field
theories, using world line path integrals (WOLPI) is possible.  References to
such a representation have appeared sporadically in the literature.  De
Carvalho, Caracciolo, and Fr\"ohlich showed that the $\phi^4$ theory in the
$N\rightarrow0$ limit is equivalent to a self-avoiding random
walk.~\cite{frohlich} These authors used the WOLPI representation to
investigate
the question of triviality in scalar field theories.  Pisarski considered a
theory of paths that, in addition to the standard arc length term also
contained
a reparametrization invariant curvature term.~\cite{pisarski} He showed
that the
coupling of the curvature term is asymptotically free. Baig, Clua, and
Jaramillo~\cite{baig} simulated a world line using Pisarski's action.  They
found a non-singular transition from a random phase to a "rigid" phase of the
world line.

  In recent years
Strassler used WOLPI representations to calculate complicated vertices in the
loop expansion of gauge theories.~\cite{strassler}  Awada and Zoller modified
Pisarski's action by adding electromagnetic interactions.~\cite{awada}  These
authors investigated the possibility of existence of a phase transition to
a new
phase of electromagnetic interactions.  The nature of their "rigid" phase
was not
completely clarified.

In the current paper we will perform a simulation of a world line in the
presence  of the curvature term and electromagnetic interactions.  We will
find a phase transition that is equivalent to the Higgs phase transition in the
field theoretic representation.  Even though no scalar self-interactions are
included yet the combination of the curvature term and the electromagnetic
interaction generates a self interaction of scalars that induces the
transition.  In the current paper  we study the system in the quenched
approximation, or, in other words, in the limit of $N\rightarrow0.$

The theory we investigate is completely equivalent to a theory of polymers with
a particular long range interaction and a stiffness term.  As the stiffness
increases there is a phase transition to a symmetry breaking phase.  In this
phase Goldstone bosons, as tadpoles, appear on propagating charge particle
lines. As Goldstone bosons appear in all theories with broken continuous
symmetries it is quite possible that they have a significance in polymer
theories as well.

Before turning to discussing the details of our model we note that
simulations in
the WOLPI representation have a considerable advantage compared to simulations
in the field theoretic functional integral representation, especially at a
large
number of dimensions, $D$.  While the number of computations in field theoretic
simulations increases as $L^D$, where $L$ is the linear size
of the system, even if we do not consider critical slowing
down or nonlocal Lagrangians (such as determinants).
The required time of simulations in the WOLPI representation increases with the
size of the system and the number of dimensions much slower, rather like $L^2D$.  
This is a considerable advantage if $D\geq 3$.

In what follows, we will investigate the Abelian Higgs theory in WOLPI
representation.  In the next section we derive the general form of the theory
and motivate the introduction of the curvature term. In Sec.~3 we discuss the
results of our Monte Carlo simulations, while in the last section we analyze
our findings and discuss possible directions for further investigations.
 \section{Higgs theory in the WOLPI representation}
 We will consider an Abelian Higgs theory with $N$ flavors.  First we will
reformulate this theory in the language of world line path integrals.  Then
we will discuss the possibility of generalization in the WOLPI representation
by introducing a curvature term.

The gauged Higgs action has the following form in the standard field theoretic
representation:
 \begin{equation}
S=S_0+S_{\rm e.m.}+\frac{g}{2}\int d^Dx [\phi^\dagger
(x) \phi(x)]^2
\label{action1}
\end{equation}
where
\begin{equation}
S_0=\int d^Dx\left\{-D_\mu\phi^{\dagger}(x)D_\mu\phi(x)+ m^2 \phi^\dagger
(x) \phi(x)\right\}.
\label{s0}
\end{equation}
and
\begin{equation}
S_{\rm e.m.} = \frac{1}{4}\int d^Dx F_{\mu\nu}(x)F_{\mu\nu}(x).
\label{sem}
\end{equation}
Here, as usual, the covariant derivative, $D_\mu$, is defined as $D_\mu=
\partial_\mu -ieA_\mu(x).$  The electromagnetic field tensor is
$F_{\mu\nu}(x)=\partial_\mu A_\nu(x)-\partial_\nu A_\mu(x).$  The complex
scalar
field, $\phi$, has $N$ flavors.  Here, and in what follows, flavor indices are
suppressed.

Our first step is to integrate out the scalar field from the
theory.  This can be done after introducing the source, $j(x)$, for the
composite operator $\phi^\dagger (x) \phi(x)$.  Using such a source the
partition function can be written as  \begin{equation} Z=\exp\left\{
\frac{g}{2}
\int d^Dy \frac{\delta^2}{\delta j(y)^2}\right\}\int d\phi(x)dA_\mu(x)\exp
\left\{ iS_0+iS_{\rm e.m.} + i\int d^Dx j(x) \phi^\dagger (x) \phi(x)\right\}
\label{part} \end{equation} The above transformation leads to a Gaussian
form  in
the scalar field.  At this moment this procedure seems to be fairly formal, due
to the complicated dependence of the action on the source function, $j(x)$.
Soon
we will see, however, that the theory can be transformed to a more manageable
form.

After the Gaussian integration over the scalar field we obtain
\begin{equation} Z=\exp\left\{ \frac{g}{2} \int d^Dy \frac{\delta^2}{\delta
j(y)^2}\right\}\int dA_\mu(x)\exp \left\{
\frac{1}{4}F_{\mu\nu}F_{\mu\nu}\right\}\frac{1}{ {\rm DET}^N} \label{part2}
\end{equation}  The determinant, DET=$\det\left[D_\mu^2+m^2+j(x)\right]$,  is
dependent on the external fields $A_\mu$ and $j$.  Its logarithm can be written
in  the path integral form~\cite{strassler} \begin{equation}
 \log{\rm DET}= -\int_0^\infty \frac{{\rm d}T}{T}\int
dx_\mu(t)\exp\left\{-\int_0^T d\tau \left[\frac{x^{\prime
2}+m^2}{2}-iex^{\prime
}\cdot A[x(\tau)]+j[(x(\tau)]\right]\right\}, \label{deter}
\end{equation}
where the boundary conditions for the quantum mechanical path integral are
$x(T)=x(0)$ and the integration is to be performed over all possible path with
the above boundary conditions.

The propagator of a scalar particle can
also be written in a functional integral form of the above type.  We obtain
\begin{equation}
\Delta(x_1-x_2)=\exp\left\{ \frac{g}{2} \int d^Dy
\frac{\delta^2}{\delta j(y)^2}\right\}\int
dA_\mu(x)\exp \left\{
\frac{1}{4}F_{\mu\nu}F_{\mu\nu}\right\}\frac{\Delta_{A,j}(x^{(1)},x^{(2)})}{
{\rm DET}^N},
\label{prop4}
\end{equation}
where
\begin{equation}
\Delta_{A,j}(x^{(1)},x^{(2)})= \int_0^\infty {\rm d}T\int
dx_\mu(t)\exp\left\{-\int_0^T d\tau
\left[\frac{x^{\prime 2}+m^2}{2}-iex^{\prime }\cdot
A[x(\tau)]+j[(x(\tau)]\right]\right\},
\label{prop1}
\end{equation}
The propagator depends on background fields $A_\mu(x)$ and
$j(x)$. In (\ref{prop4}) the boundary conditions  are $x(0)=x^{(1)}$ and
$x(T)=x^{(2)}$.

The expression for the partition function and the propagator can be further
simplified and the formal functional differentiations with respect to source
$j(x)$ and functional integration over the electromagnetic field can be
calculated.  This can be done after expanding $DET^N$ in a power series
of $N$. Then the derivative, $\int d^Dy\delta^2/\delta
j(y)^2$ simply generates a contact term of
$\int_0^{T_i}d\tau_i\int_0^{T_j}d\tau_j\delta^D[x(\tau_i)-x(\tau_j)]$ in the
action.  After
integrating over the electromagnetic field one obtains the following
representation in Euclidean metric
\begin{eqnarray} Z &= &\sum_{n=0}^\infty
\frac{N^n}{n!}\int_0^\infty \prod_{i=1}^n \int_0^\infty\frac{dT_i}{T_i}\int
dx_\mu^{(i) }(\tau_i) \exp\Bigg\{ -\sum_{i=1}^n
\int_0^{T_i}d\tau_i\frac{x^{(i)'2}+m^2}{2}\nonumber \\
&-&\frac{e^2}{2}\sum_{i,j}^n\int d\tau_i d\tau_j x^{(i)'}_\mu(\tau_i)
x_\nu^{(j)'}(\tau_j)D_{\mu\nu}\left(x^{(i)}(\tau_i)-x^{(j)}(\tau_j)\right)
\nonumber\\ &-& \frac{g}{2}\sum_{i,j}^n\int d\tau_i d\tau_j
\delta\left(x^{(i)}(\tau_i)-x^{(j)}(\tau_j)\right)\Bigg\},
\label{action3}
\end{eqnarray}
where $D_{\mu\nu}(x-y)$ is the $D$-dimensional photon
propagator.  The physical interpretation of this form is as follows. The
theory describes the dynamics of closed interacting world lines such that the
electromagnetic interaction appears as a long range current-current type
interaction between bits of the world line.  The scalar self-interaction
results
in a contact term. This is also a long range
interaction in polymer physics because faraway segments along the polymer
interact with each other.  At infinite coupling (Ising model) this contact term
will generate a self-avoiding walk. Note that both types of path-path
interactions appear as interactions between different paths and as
self-interactions of paths, as well.  Our model is completely equivalent to a
polymer, albeit with the unusual electromagnetic (current-current)
interaction.

 The term resulting from the
scalar self interaction is well known in polymer physics.~\cite{book} It can be
 rewritten as
\begin{equation}
S_{\rm scalar} = \frac{g}{2}\int d{\bf r} \, [c({\bf r})]^2,
\label{scalar-term}
\end{equation}
where $c({\bf r})$ is the polymer density function
\begin{equation}
c({\bf r}) = \int d\tau \, \delta[{\bf r - x}(\tau)].
\label{density}
\end{equation}
In polymer physics (\ref{scalar-term}) is the first nontrivial term of a
virial-expansion in terms of the polymer density.~\cite{book}

An expression for the scalar propagator can be found in a manner, very
similar to that of the partition function.  Using (\ref{prop4}) and
(\ref{prop1}). after manipulations, similar to the ones used for the
calculation
of the partition function, the propagator can be brought to a form, similar to
(\ref{action3}).  We only present the form of the propagator in the limit
$N\rightarrow0$
as in the current paper we will not go beyond the quenched approximation.  We
obtain \begin{eqnarray} \Delta(x_1-x_2)&= &\int_0^\infty
dT\int dx_\mu(\tau)
\exp\Bigg\{ -
\int_0^{T}d\tau\frac{x^{'2}+m^2}{2}-\frac{1}{2}\int_0^T d\tau_1 d\tau_2
\nonumber \\ & \times & \left[ e^2\, x^{'}_\mu(\tau_1)
x_\nu^{'}(\tau_2)D_{\mu\nu}\left(x(\tau_1)-x(\tau_2)\right)+ g
\delta\left(x(\tau_1)-x(\tau_2)\right)\right]\Bigg\},
\label{prop3}
\end{eqnarray}
where the world line $x_\mu(\tau)$ satisfies the constraint $x(0)=x_1$, and
$x(T)=x_2$.

The expression (\ref{prop3}) contains divergences.  These divergences
partly come
from the delta function of the contact density self-interaction.  That term can
be regularized by smearing the delta function to a ball of radius $a$.
Similarly, the electromagnetic propagator appearing in the action becomes
singular at short distances.  Awada and Zoller recommended a
regularization by separating the paths for $x(\tau_1)$ and $x(\tau_2)$ in the
electromagnetic interaction term.~\cite{awada}. Such a regularization is gauge
invariant. If  the two trajectories are shifted away from one another along the
principal normal then one can show that one
introduces a finite renormalization of the form $\pm e^2\int d\tau/R$, where R
is the radius of curvature.  In fact, if we do not insist on splitting the
curves along the principal normal, but say at a fixed angle from it, then
we can get an arbitrary finite
contribution of the form   \begin{equation} \delta S_{\rm
curv} = \chi \int \frac{d\tau}{R},
\label{curv}
\end{equation}
where $\chi$ is an
arbitrary constant satisfying $|\chi|<e^2$.  This is exactly the type of term
proposed by Pisarski~\cite{pisarski}. This term is reparametrization and scale
invariant.  Since certain regularization procedures require its presence
(albeit with a fixed regularization dependent finite coefficient) there is no
reason why it would be forbidden in the world line representation.  As derived
from the field theoretic representation its value is finite, fixed, and
regularization dependent.  In what follows we will include such a term in the
action, with an arbitrary coefficient, and prove that its presence induces a
phase transition in Abelian Higgs theories.  Thus, the theory we study is
 a generalization of the standard Higgs Lagrangian, but one that is
admissible in the WOLPI representation.

The reparametrization invariance can be used to fix the world line such that
$|x'(\tau)|=1$. We will use this gauge in the rest of this paper.  We will
also introduce the effective mass, $M=(x^{'2}+m^2)/2$.

\section{Simulations}
To perform simulations we are required to discretize the world line, replacing
it by a finite number of straight segments. Unlike in standard simulations, in
which fields are defined at discretized points of the spacetime, it is not
necessary to tie WOLPI representation simulations to a lattice. In fact,
there is a definite advantage of allowing the vertices of the world line to
move
to arbitrary points of ${\cal R}^D$.  The critical region of our investigations
will be at large values of the curvature term and of the electromagnetic
coupling.  At these values the curvature of the world line is small and the
lattice approximation that allows the change of direction of the world line
only by $\pi/2$ is very crude. Fortunately, there is no real
numerical advantage of working on a lattice either.  The most computer
intensive
part of evaluating the action is the calculation of the electromagnetic energy
that is not very strongly dependent on whether the vertices are on ${\cal Z}^D$
or on ${\cal R}^D$.

There is another, even more important, advantage of not doing the
simulations on
a lattice.  As (\ref{prop3}) shows the calculation of the propagator requires
integration over the length of the world line.  On a lattice, with a fixed
number of segments, the length of the world line is fixed. Our Monte Carlo
procedure that allows points on the world line move freely automatically
averages over the length of the world line to some extent.

As we mentioned earlier, there is a significant
advantage of world line simulations compared to simulations involving fields.
When one simulates a local field theory on a lattice, even if critical slowing
down effects, getting more and more significant at large lattice sizes, are not
considered, the number of floating point operations per update increases as a
multiple of $L^D$, where $L$ is the linear size of the lattice and $D$ is the
number of spacetime dimensions.  World line simulations with long range forces
(such as an electromagnetic interaction) require $L^2D$ elementary calculations
(update $L$ points which all interact with all other $L$ points).  Thus, when
$D\geq3$ there is a definite advantage of performing simulations in the world
line representation.

The discretized form of the action, appropriate for numerical simulations, is
as follows:
\begin{equation}
Z = \int \prod_{i=1}^L d{\bf r}^{(i)} \exp\left\{ -  \sum_{i=1}^L [
M\,|\Delta {\bf r}^{(i)}| + \chi\sin(\theta_{i,i+1}/2)]
-\frac{e^2}{2}\sum_{i,j}^L  \frac{\Delta {\bf r}^{(i)} \cdot \Delta{\bf
r}^{(j)}}{({\bf r}^{(i)}-{\bf r}^{(j)})^2+a^2}\right\}, \label{discrete-action}
\end{equation}
where
$\theta_{i,i+1}$ is the angle between the $i$th and $(i+1)$st segments of the
world line, and $\Delta{\bf r}^{(i)}={\bf r}^{(i+1)}-{\bf r}^{(i)}$.  We
use here the Feynman gauge for the electromagnetic propagator. The cutoff
distance, $a$, sets the length scale and as such it can be chosen
arbitrarily to
be 1. Then the effective mass, $M$, is chosen to be low enough,
corresponding to
a small value of $a$.  In most of the simulations we used $M=0.2$. The
self-interaction term of the scalars is not expected to affect  the phase
transition significantly because the combination of the electromagnetic
term and
of the curvature term generate an effective scalar self-interaction anyway.
Therefore, in the current paper, we set $g=0$ and defer the investigation
of the
case of $g>0$ to a future publication.

In our simulations a sweep constituted of changing the coordinates of all $L$
vertices, one by one. We restricted these changes within a fixed
radius and used the Monte Carlo method to accept or reject individual moves.
The radius of allowed moves was varied to achieve a 50\% acceptance ratio.

We performed simulations both at $D=3$ and $D=4$. We also used two different
forms for the electromagnetic propagator.  First we used both at $D=3$ and at
$D=4$ the $D=4$ inspired form of $D_{\mu\nu}(x) =\delta_{\mu\nu}
/(x^2+a^2)$, as given in (\ref{discrete-action}). The special property of this
propagator that it is scale invariant in every dimension, at least at
vanishing cutoff parameter, $a$.  Then at $D=3$ we also tried the physically
correct cutoff electromagnetic propagator, $D_{\mu\nu}(x) =\delta_{\mu\nu}
/\sqrt{x^2+a^2}$.  Finally, we simulated both closed loops and open world
lines.
The phase structure of the system was largely independent of the
choice of dimension or that of the electromagnetic interaction. This fact
also points into the direction that the transition is related to the breaking
of the same symmetry, charge conservation, irrespective of the details of the
interaction.

We simulated both closed and open world lines.  Closed world lines
correspond to the condensate or propagator corrections on the gauge boson,
while
open world lines simulate the propagator of charged particles.  Rather than
fixing the endpoints of the world line at $x_1$ and $x_2$, we allowed the
endpoints of open world lines to move freely.  This allowed us to measure the
probability distribution of $R$, the end-to-end distance of the world line,
which is proportional to the propagator.  This distribution plays a significant
role in the analysis of the nature of the phase transition.
\begin{figure}[t]
\epsfxsize=9cm
\epsfysize=6cm
\centerline{
\epsfbox{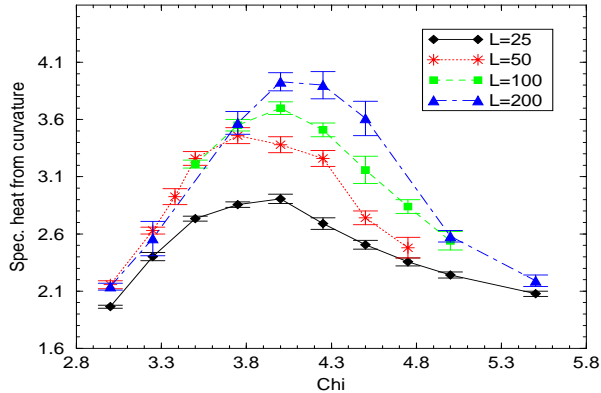}}
\caption{\em The dependence of the specific heat, $\partial^2 Z/\partial\chi^2$
on $\chi$ in the critical region at fixed $e^2=0.5$. The purpose of the connecting 
lines is to guide the eye.}
\label{fig1}
\end{figure}
The salient feature of simulations is the developing of a significant maximum
in the
specific heat, $d^2Z/d\chi^2$, as a function of $\chi$ and fixed $e$, provided
the fixed value of $e$ is sufficiently large.  The peak also appears in
$d^2Z/de^2$ at fixed, sufficiently large $\chi$ and increasing e. The
height of the peak increases rapidly as a function of the number of segments on
the world line, $L$, which controls the size of the system. This is a telltale
sign of a second order phase transition.  Fig.~\ref{fig1} shows the dependence of
specific heat of the curvature term
 on $\chi$ at fixed $e^2=0.5$ and $L=$25, 50, 100, and 200 in the critical
region.  There is a clear evidence for a second order phase transition at
around
$\chi_c=4$. Note the the difference between the curve at $L=25$ on one hand and
the curves at larger $L$ on the other hand.  This difference suggests that at
$L=25$ large $L$ asymptotic scaling laws, used to extract critical
behavior, are
not always applicable.

\begin{figure}[t]
\epsfxsize=9cm
\epsfysize=6cm
\centerline{
\epsfbox{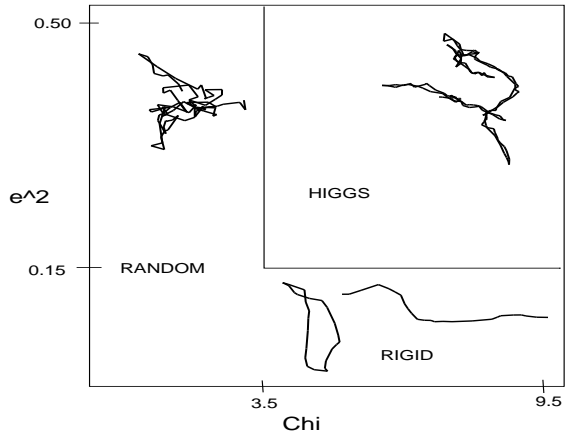}}
\caption{\em The phase diagram of the system in the $\chi-e^2$ plane. Characteristic
shapes of world lines are shown in various regions.}
\label{fig2}
\end{figure}

By changing $e^2$, as well, and by exchanging the role of $\chi$ and $e^2$ one
can arrive at the phase diagram sketched in Fig.~\ref{fig2}.  It is interesting to note
that the phase transition occurs always at non-vanishing value of both
$\chi$ and
of $e^2$.  Thus, the smooth transition observed by Baig et
al.~\cite{baig} in the absence of electromagnetic interactions is distinct from
the phase transition we observe.

To understand the nature of this phase transition it is very
instructive to observe the characteristic shapes of world lines
in various regions of the phase diagram. These shapes are shown in the
appropriate regions of the phase diagram in Fig.~\ref{fig2}. Below the phase
transition (low $\chi$) the world line strongly resembles a random walk.   In
fact, it is easy to prove that the world line must follow a random walk at
vanishing $e^2$ no matter how large the curvature term is.This is true for
every
polymer with an action containing finite range correlations only.~\cite{book}
At low $e^2$ and high $\chi$ it does not look like a random walk only
because of
limitations in the length of the path as the effective radius of the random
walk
becomes larger than $L$.   Above the phase transition segments of the world
line
running in opposite directions pair up to form tree diagrams of a theory with
paired double lines.  This signals the appearance of bound states of zero mass
(vanishing total four momentum) and the same quantum number as the charge
operator, formed from bilinears of the charged scalars.  This is clearly a
Goldstone boson.  As the Goldstone boson makes its appearance at the phase
transition, the phase transition should be a Higgs-type transition.

Before we were
able to check our hypothesis concerning the nature of the phase transition we
needed to solve a crucial problem affecting simulations near or above the
critical line.  By employing our local update procedure  the
convergence of simulations for the specific heat became intolerably slow at
$\chi\sim\chi_c$ or $\chi>\chi_c$.  This phenomenon is very reminiscent to the
one observed in simulations of random surfaces by Ambjorn, Bialas, Jurkiewicz,
Burda, and Petersson~\cite{ambjorn}  who observed that above a critical point
baby universes form and are very stable against local changes.  They would not
change over enormous computer simulation times. The use of global changes,
amputation and reattachment of baby universes was required to speed up the
algorithm.

 We are faced with a very similar problem in our world line simulations.  The
limbs of tree diagrams formed from particle-antiparticle bound states are
extremely stable against local changes.  Therefore, we introduced two global
moves that can change these configurations effectively.  Both of these moves
intend to change configurations with paired world lines. Therefore, after every
sweep of local updates, we employ  these moves to a randomly chosen pair of
 points, that are far away from each other along the world line, but are
closer than a pre-determined distance in the ${\cal R}^D$ target space.

The global
moves we employ are: 1) bending the world line along the connecting line of two
points that are close in ${\cal R}^D$ space but widely separated along the
world line, 2) amputating the part of the world line beyond a similarly
selected pair of points and attaching the amputated part to an other pair of
randomly selected similarly close pair of points.  These moves satisfy the
detailed balance requirement and were constructed in a way that they would not
change the action by a large amount so that their acception rate would not be
intolerably low.  The result of the introduction of these global moves was a
substantial increase in the speed of convergence of simulations near or above
the phase transition.  Without the introduction of global moves it is
practically
impossible to perform simulations at a world line length $L\geq100$. It is
quite
possible that other, more sophisticated global moves would further speed up the
the algorithm.

Though the shapes of the world line in the two phases shown in Fig.~\ref{fig2} give a
telltale sign of the nature of the phase transition, we need to find
quantitative support for our hypothesis of symmetry breaking transition.  We
found that the best measure describing the nature of the strong coupling phase
is the end-to-end correlation function of open world lines.  As we mentioned
before, when we simulate open world lines then instead of fixing the
endpoints to
calculate the scalar propagator between a pair of points we allow the endpoint
to move freely.  By this we are able to measure the relative probability of
propagation to points at varying distance.

The distribution of the end-to-end  distance at $e=0$  is
Gaussian.  This is true for the distribution of random walks with arbitrary
finite range correlations in the action.~\cite{book} For such random walk like
paths one has the following form for the distribution of the end-to-end
distance, $R$  \begin{equation} P(R) \sim
\exp\left\{ - \frac{3R^2}{2b^2 L}\right\}, \label{endtoend} \end{equation}
where
$b$ is the effective bond length. At $e^2=0$ there is only a ``nearest
neighbor''  correlation along the world line. At finite values of $e^2$ this
distribution is expected to be renormalized and the $L$ factor in the
denominator of the exponent of (\ref{endtoend}) replaced by a factor of
$L^{2\nu}$, where $\nu$ is the correlation length exponent.
On the contrary, in a phase dominated by a bound state, one expects the
end-to-end distance distribution to be independent of the length of the world
line.  The segment of the world line between the two end points corresponds to
the propagation of the charged scalar particles. The finiteness of this segment
proves a bound system.  This can be contrasted to the doubled up
part of the world line that corresponds to the propagation of massless bound
state formed from the scalars and propagates freely to a distance of
$O(L)$.
In particular, if the potential creating the bound state is bounded, then the
single particle propagator decreases exponentially at large distances, such as
 \begin{equation} P(R) \sim \exp\left\{ - M R\right\},
\label{endtoend2} \end{equation}
where $M$ is the energy gap.  
For an unbounded, confining potential the propagator decreases faster for large
distances. In that case, the decrease of the propagator at large distances is
faster than exponential.

\begin{figure}[t]
\epsfxsize=9cm
\epsfysize=5.5cm
\centerline{
\epsfbox{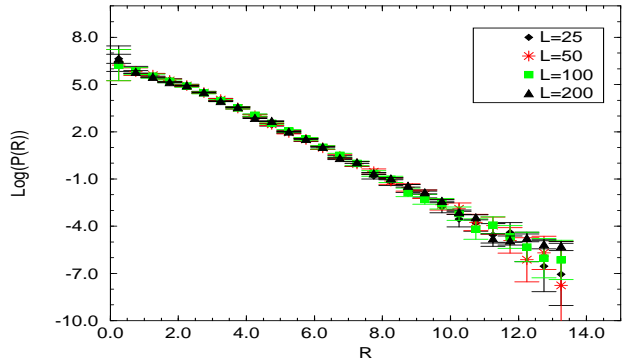}}
\caption{\em The dependence of the propagator, $P(R)$, on $R$ at $\chi=5.0$
and $e^2=0.5$, in the broken phase. A straight line corresponding to an exponential
dependence of P(R) on R gives a perfect fit at $R>2$.}
\label{fig3}
\end{figure}

\begin{figure}[hbt]
\epsfxsize=9cm
\epsfysize=5.5cm
\centerline{
\epsfbox{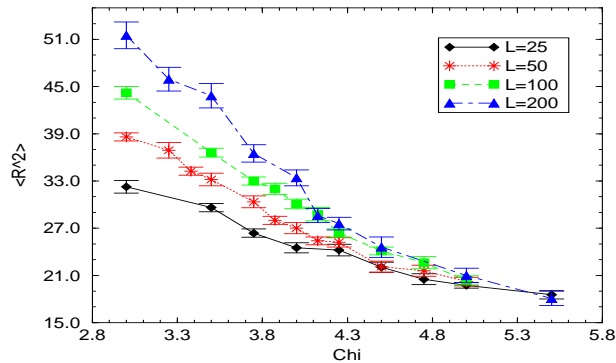}}
\caption{\em The square of the end-to-end distance as a function of $\chi$
in the critical region at fixed $e^2=0.5$. The purpose of the connecting lines 
is to guide the eye.}
\label{fig4}
\end{figure}

\begin{figure}[hbt]
\epsfxsize=9cm
\epsfysize=6cm
\centerline{
\epsfbox{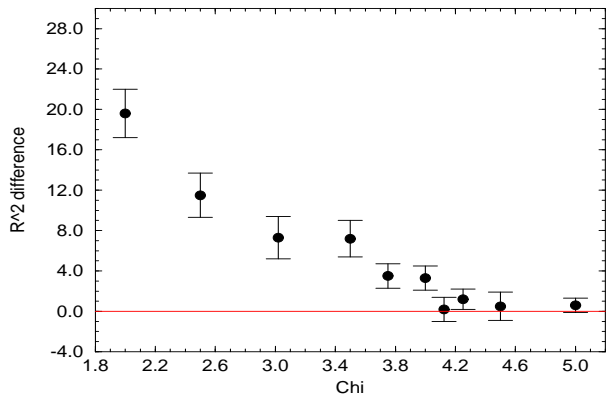}}
\caption{\em The difference of the squares of the end-to-end distances
at $L=200$ and $L=100$ as a function of $\chi$.}
\label{fig5}
\end{figure}

Fig.~\ref{fig3}. shows the dependence of the propagator, $P(R)$, on $R$ at $\chi=5.0$
and $e^2=0.5$, in the broken phase. The distributions for $L=25$, 50, 100, and
200 perfectly track each other and at distances $R>2$ are in complete
agreement with an exponential dependence on $R$ (a straight line in the figure,
having a logarithmic scale).
The transition between the two regimes is illustrated on the
plot, in Fig.~\ref{fig4}, of $\langle R^2\rangle$, where $R$ is the end-to-end distance,
for $L=25$, 50, 100, and 200, as a function of $\chi$, at fixed $e^2=0.5$.
Clearly, below the phase transition ($\chi_c\simeq4.2$) there is a strong
dependence on $L$, that can be well fitted to the formula
\begin{equation}\langle
R^2\rangle\sim A+ B\ L^{2\nu}
\label{averageetoe}
\end{equation} 
that follows from a renormalized version of (\ref{endtoend}),
while above the phase transition $\langle R^2\rangle$ is smaller and
independent
of $L$. Omitting data at $L=25$ the fit to
(\ref{averageetoe}), assuming the mean field (free) value, $\nu=1/2$, at
every value of $\chi< 4.25$, is perfect. Using the dependence of the
coefficient $B$ in (\ref{averageetoe}) using a linear extrapolation, we can
extrapolate to a critical point at $\chi_{\rm crit}=4.4\pm0.3$.  Unfortunately,
the closer one goes to the critical point the less data at smaller values of
$L$ fit the asymptotic formula (\ref{averageetoe}). Therefore, a better
estimate for the critical point can be obtained if we use only data at the
highest values of $L$, $L=200$ and $L=100$.  The difference of $\langle
R^2\rangle$ is plotted for these values in Fig.~\ref{fig5}. These data do not allow for
any meaningful fit of the critical behavior, but provide a rough estimate of
the critical point as $\chi_{\rm crit}=4.2\pm0.4$.

In a similar manner we measured the propagator of the particle-antiparticle
system.  Since particle-antiparticle propagators are non-vanishing to distances
of $O(L)$ the normalization of these propagators depends on the value of
$L$. 
\begin{figure}[t]
\epsfxsize=9cm
\epsfysize=6cm
\centerline{
\epsfbox{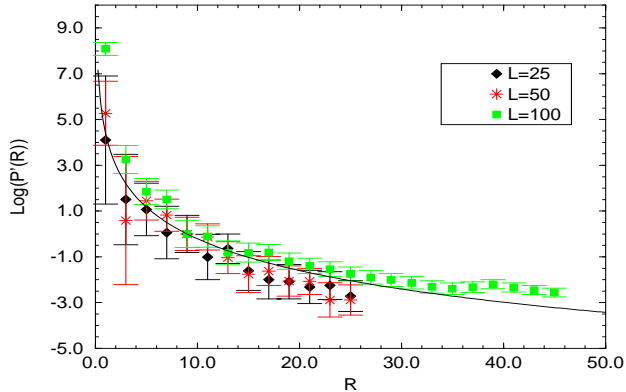}}
\caption{\em The dependence of the propagator of the particle-antiparticle system,
$P'(R)$, on $R$ at $\chi=5.0$ and $e^2=0.5$, in the broken phase. As these 
propagators have an infinite range and extend to $R\sim L$, they are normalized 
to equal values at $R=9$. The curve is $-2 \log(R)$, corresponding to a massless 
free particle.}
\label{fig6}
\end{figure}
Therefore, we plot these propagators in Fig.~\ref{fig6} using a normalization so
that they coincide at the arbitrarily chosen point, $R=9$.  
Using this normalization the propagators show a remarkable agreement with each other and
with the curve $P'(r)\simeq 1/r^2$, the true scalar propagator (up to
logarithmic corrections) of a zero mass particle in four dimensions.
These particle-antiparticle propagators can be obtained  either from open or
from closed loops. Closed loops also correspond to corrections to the gauge
boson propagator. Having a zero mass pole at $N>0$ they would generate a
non-vanishing mass to the gauge propagator in agreement with the Higgs
mechanism.

There are various other ways to determine the critical properties of the
system.
One is using finite size scaling.  The behavior of specific heat near the
phase transition is~\cite{fisher}
\begin{equation}
C(\chi) = A + B \left\{[\chi-\chi_{\rm max}(L)]^2+ F
L^{-2/\nu}\right\}^{-\alpha/2}, \label{spec-heat}
\end{equation}
where $\alpha$ is the specific heat exponent, $\nu$ is the correlation length
exponent, and $\chi_c(N)$ is the maximum of the specific heat curve and it
depends on the size of the system as
\begin{equation}
\chi_{\rm max}(L) = \chi_c - c\, L^{-1/\nu}.
\label{crit-point}
\end{equation}
Alternatively, if the critical exponent $\alpha$ vanishes, as it does in the
mean field approximation, then the singularity of the specific heat is
logarithmic, as in
\begin{equation}
C_\chi=A + B \log\left\{[\chi-\chi_{\rm max}(L)]^2+ F L^{-2/\nu}\right\}.
\label{spec-heat2}
\end{equation}
Our fits to the specific heat are consistent with the mean field
prediction of a logarithmic singularity of the specific heat, but a small value
of the exponent $\alpha$ cannot be excluded.  Our data are insufficient to make
any meaningful prediction for the correction term, $F L^{-2/\nu}$.  We could,
however, analyze the dependence of $\chi_{\rm max}(L)$ on $L$ according to
(\ref{crit-point}). Using $\nu=1$, we found that $\chi_{\rm crit}=4.15\pm0.08$,
in agreement with our earlier estimate of the critical point from the
end-to-end distance distribution.

Our understanding of the world line configurations of
Fig.~\ref{fig2} is now the following: In the Higgs phase corrections to the
Higgs propagator contain tadpole diagrams.  These tadpole diagrams are
generated
by the interaction of Higgs particles with Goldstone bosons and Goldstone
bosons
with each other. As we simulate a single world line only (quenched
approximation)  only  tree diagram tadpoles  formed from the Goldstone boson
propagator appear. We expect that in the approximation, $g>0$,  loop diagrams
formed from the Goldstone boson would also make their appearances.  The
expected
effect of having a finite number of flavors, $N$, will be discussed in the next
section.

\section{Conclusions} In the world line path integral
representation the Abelian Higgs theory admits the addition of a
curvature term to the Hamiltonian.  We showed that the addition of such a term
with a sufficiently large coupling generates a second order phase transition
from the symmetric phase to a broken phase in which particle-antiparticle
bound states can propagate to large distances or, in other words, form zero
mass
(Goldstone) bosons. The critical properties of the system were
investigated and were found, with admittedly large errors, to be in agreement
with a mean field behavior.   As the curvature term appears in a natural manner
in the world line representation (certain ultraviolet regularizations
require its
appearance) we expect that this theory is in the same universality class as the
Higgs theory in the field theoretic representation, in which there is no
term in
the Lagrangian, equivalent to the curvature term.  This identification of the
universality classes leads, however, to  several natural questions that need
be answered to fully understand the significance of our results:
\begin{enumerate} \item In four dimensions the Higgs phase transition should be
of first order, as it was shown a long time ago by Coleman and
Weinberg.~\cite{coleman} Why do we find a second order phase transition in our
simulations?
  \item In a
Higgs theory the charge symmetry is broken. Thus, the single charged
particle and
particle-antiparticle sectors should mix with each other.  Yet we find separate
asymptotic states in these sectors.
 \item  In a Higgs theory
no zero mass particles should exist, as the Goldstone boson is absorbed by
the longitudinal component of the massive gauge boson.  What is the reason than
that we still seem to see a zero mass asymptotic state? \end{enumerate}

The short answer to questions 1) and 3) is that the discrepancy is the
consequence of the quenched approximation that we use in the current paper.  In
the $N\rightarrow0$ limit closed scalar loops do not make their appearances.
These scalar loops are needed to generate a mixing between
particle-antiparticle
states and gauge bosons.  At $N>0$ closed scalar loops appear.  In a simulation
they should be produced freely, as states with a finite number of loops give
a finite contribution to the partition function.  In paired up lines,
formed from
particle-antiparticle states, the line pair would go over into a pure gauge
state with high probability.  This is tantamount to spontaneously forming
closed
bosonic loops.\footnote{ The situation is reminiscent to the problem one faces
when investigating confinement in lattice QCD with colored matter fields
(quarks). As quarks are separated to a large distance quark-antiquark pairs
(mesons) are produced from the vacuum allowing the separation of the original
quarks.} Particle-antiparticle systems then cannot propagate freely any more,
but after a finite distance they annihilate and form, along with virtual gauge
boson states, the massive longitudinal component of the physical gauge boson.
Our conclusion is then that the $N\rightarrow0$ limit is a {\em Goldstone limit}
of the Higgs theory. The theory should then be in the universality class of
a pure scalar theory. That theory has a second order phase transition with mean
field exponents.

The second question in the above list concerns the mixing of states with
different particle numbers.  Note that even in the field representation such a
mixing occurs only if we shift the fields and expand around the correct
vacuum.
In the absence of such a shift the Goldstone boson can only appear in the $Q=0$
sector, i.e. in the sector of the broken generator. While expanding around the
false vacuum the charge operator, $Q$, seems to be conserved.  In the world
line
representation the value of the charge operator on a surface equals the
number of
world lines crossing the surface.  Clearly, unless we allow the creation of
world lines from vacuum, charge is formally conserved. Putting it slightly
differently, if we use eigenstates of the charge operator,  such as the states
built from a definite number of world lines, then we do not see the breaking of
charge symmetry, but we still see the appearance of Goldstone bosons signaling
the phase transition.

Another way to argue for the second order nature of the phase transition in the
$N\rightarrow0$ limit is as follows.  If the electromagnetic interactions
of the
single world line are sufficiently strong they provide a scalar self-interaction
that breaks the symmetry.  The first order transition in the Coleman-Weinberg
mechanism is generated by gauge loops. The logarithmic divergence of the
effective potential in one loop order comes from the two gauge boson
intermediate state diagram in particle-antiparticle scattering.~\cite{coleman}
In the absence of this diagram and similar diagrams the Coleman-Weinberg
mechanism does not operate.

We can
only speculate why the nonzero value of the curvature term is required for
the phase transition.  Clearly, in forming the zero mass bound states the
electromagnetic energy competes with the entropy, as the particle and
antiparticle have much larger phase space if they are not tied together into a
bound state. Apparently, in the absence of the curvature term the
electromagnetic
energy can  never win this competition.  The introduction of the curvature term
effectively decreases the entropy as world lines become straight. The
straight particle and the antiparticle world lines only need to be made parallel
(involving a finite number of degrees of freedom) to form the bound state.
This
is a competition that the electromagnetic energy can easily win.

 It is clear from the above discussion that it would be interesting to perform
simulations at $N>0$, where  the phase transition should switch to first order.
 We conjecture that the $g\rightarrow0$ limit does not influence the nature
of the transition.  This conjecture should also be checked against simulations.
In the future we intend to perform appropriate simulations to investigate these
effects.

\noindent {\bf Acknowledgments:} The authors are indebted to Professor A. Baig
for discussions and for an exchange of codes.  Discussions with Drs.  I.
Shovkovy and L.C.R. Wijewardhana are also greatfully acknowledged.  This
work was
supported in part by the U.S. Department of Energy under grant DOE
DE-FG02-84ER-40153 and by a grant of the Ohio Supercomputer Center where
most of
the calculations were performed.

 \end{document}